\documentstyle[12pt,epsfig]{article}
\def\bge{\begin{equation}}
\def\ene{\end{equation}}
\def\bg{\begin{eqnarray}}
\def\en{\end{eqnarray}}
\def\nn{\nonumber}

\def\ubar{{\bar{u}}}
\def\dbar{{\bar{d}}}
\def\sbar{{\bar{s}}}
\def\cbar{{\bar{c}}}
\def\d0bar{{\bar{D}^0}}
\def\Dbar{{\bar{D}}}

\def\vr{\vec{r}}
\def\vx{\vec{x}}

\def\e{\epsilon}
\def\pomega{{\omega}}
\def\mb{\mbox}
\def\bm{\boldmath}

\begin{document}

\begin{center}
{\bf \mb{\bm$\eta$}, \mb{\bm$\omega$} 
AND CHARMED MESIC NUCLEI}
\footnote{
Invited talk at  
the XIV International Seminar on High Energy Physics Problems\\
"Relativistic Nuclear Physics and Quantum Chromodynamics",
Dubna, Aug. 17-22, 1998}

\vskip 5mm
K. Tsushima

\vskip 5mm

{\small
{\it
Special Research Center for the Subatomic Structure of Matter (CSSM)\\
and Department of Physics and Mathematical Physics\\
The University of Adelaide, SA 5005, Australia
}
\\
{\it
E-mail: ktsushim@physics.adelaide.edu.au
}}
\end{center}

\vskip 5mm

%
\vspace{-5.5cm}
\hfill ADP-98-73/T340
\vspace{5.5cm}
%

\begin{center}
\begin{minipage}{150mm}
\centerline{\bf Abstract}
Using the quark-meson coupling (QMC) model, we   
investigate theoretically whether $\eta$ and $\omega$ mesons 
form meson-nucleus bound states. 
We study several nuclei from $^{6}$He to $^{208}$Pb, including those 
which are the final nuclei in the proposed experiment  
at GSI. Results for the $\omega$ are compared 
with those of the Walecka model. Our results suggest that one should 
expect to find $\eta$- and $\omega$-nucleus bound states in all 
these nuclei. Furthermore, we investigate the possibility of 
charmed mesic nuclei. It is shown that the $D^-$ meson 
will inevitably form narrow bound states with $^{208}$Pb.
\end{minipage}
\end{center}

\vskip 10mm
\section{Introduction}

The study of the properties of hadrons in a dense and/or hot  
nuclear medium is one of the most exciting new directions
in nuclear physics. In particular, the medium modification of the
light vector ($\rho$, $\omega$ and $\phi$) meson masses 
is expected to provide us information concerning 
chiral symmetry (restoration) in a nuclear medium.

For example, the experimental data obtained at 
the CERN/SPS by the CERES~\cite{ceres} and HELIOS~\cite{hel} 
collaborations has been interpreted as evidence
for a downward shift of the $\rho$ meson mass
in dense nuclear matter~\cite{li}. Further experiments 
are planned at TJNAF~\cite{jlab} and GSI~\cite{gsi} to measure 
the dilepton spectrum from vector mesons produced in nuclei.  

Recently, a new, alternative approach to study meson mass shifts in nuclei
was suggested by Hayano {\it et al.}~\cite{hayano}, 
involving the (d, $^3$He) reaction to produce
$\eta$ and $\omega$ mesons with nearly zero recoil.
If the meson feels a large enough, attractive (Lorentz scalar) force inside
a nucleus, the meson is expected to be bound.
This may be regarded as a consequence of the 
partial restoration of chiral symmetry. 

Stimulated by this experimental proposal, theoretical investigations 
have been made by Hayano {\it et al.}~\cite{hayano2},  
Klingl {\it et al.}~\cite{kliomega} and our group~\cite{etao} 
for various $\eta$- and $\omega$-mesic nuclei, including those 
which are the final nuclei in the proposed experiment. 
Our investigations on the mesic nuclei have been based on the 
quark-meson coupling (QMC) model (and the Walecka model for 
the $\omega$). A detailed description of the QMC model which 
is relevant for this report can be found in 
Refs.~\cite{qmc} -~\cite{tony}.

The result for the $\omega$-meson is especially interesting. Because
it consists of almost pure light quark-antiquark pairs,
$q$-$\bar{q}$ ($q=u,d$),   
in an isoscalar nucleus the $q$ and $\bar{q}$ feel
equal and opposite vector potentials. 
This means that the effect of the Lorentz scalar potential 
is unmasked which leads to expectations of
quite deeply bound $\omega$-nucleus   
states~\cite{hayano} -~\cite{etao}.

Concerning charmed mesic nuclei, it is in some ways even more 
exciting, in that it promises more specific information on the
relativistic mean fields in nuclei and the nature of dynamical chiral
symmetry breaking. We focus on systems containing an anti-charm quark
and a light quark, $\bar{c}q$, which have no strong decay channels if bound.
If we assume that dynamical chiral symmetry breaking is the
same for the light quark in the charmed meson as in purely light-quark 
systems, we expect the same coupling constant,
$g^q_\sigma$, in the QMC model.

In the absence of any strong interaction, the $D^-$ will form
atomic states, bound by the Coulomb potential. The QMC model is used here 
to estimate the effect of the strong interaction. 
The resulting binding for, say, the 1s level in $^{208}$Pb
is between ten and thirty MeV and should provide a very clear
experimental signature. On the other hand, 
although we expect the D-meson (systems of $\bar{q}c$) will   
form deeply bound $D$-nucleus states, they will also couple
strongly to open channels such as $D N \rightarrow B_c (\pi's)$, with 
$B_c$ a charmed baryon. Thus, it requires an accurate calculation 
for the widths in a nucleus to estimate the  
bound state energies relevant for the experimental measurements. 
Unfortunately, because our present knowledge does not permit such 
an accurate calculation of the corresponding widths, 
results for the $D$-mesic nuclei may not give useful information  
for experimenters.  

\section{Meson masses in nuclear matter and finite nuclei}

\begin{figure}[hbt]
\begin{center}
\hspace*{-1cm}
\epsfig{file=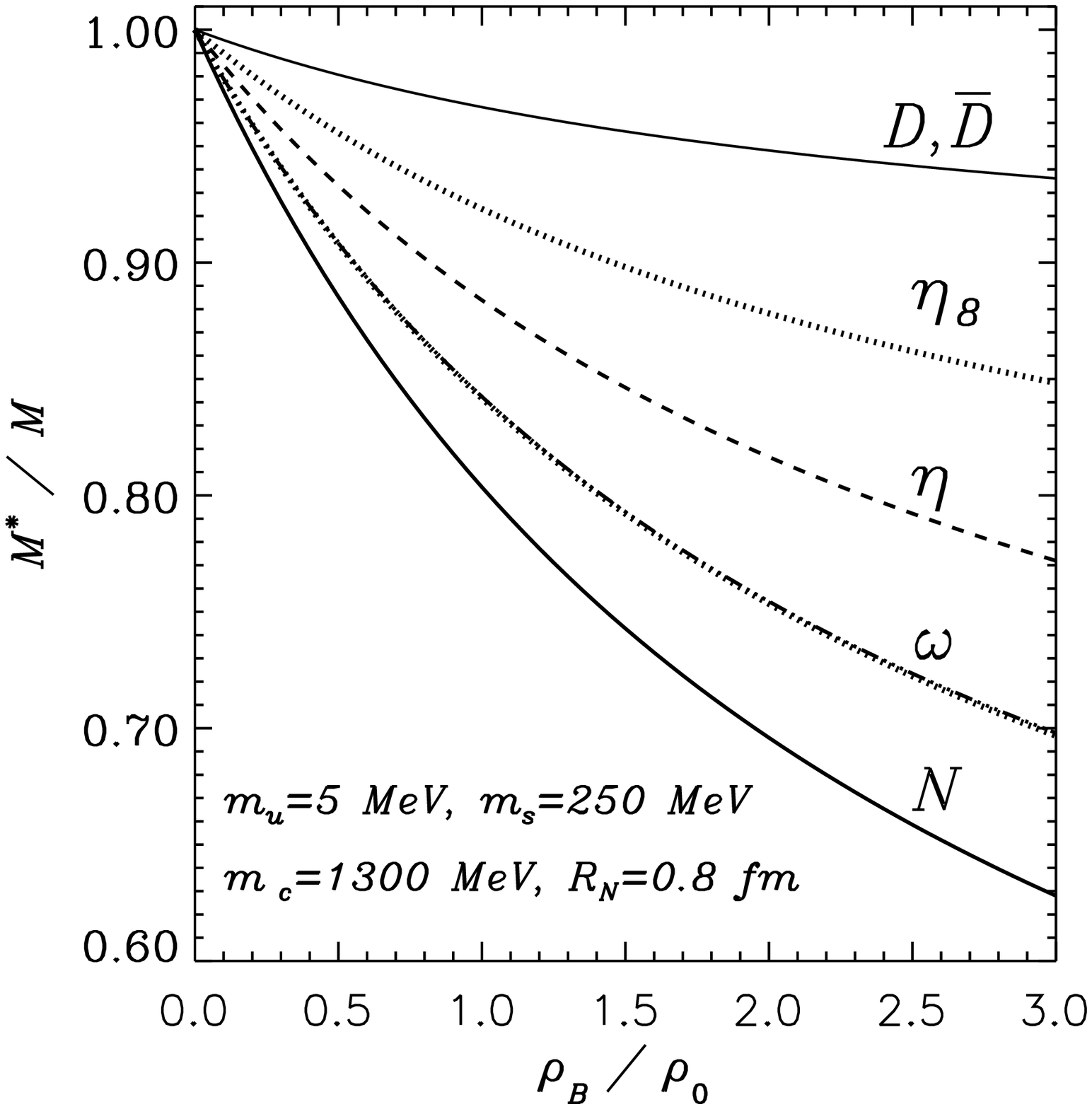,height=7cm}\quad
\epsfig{file=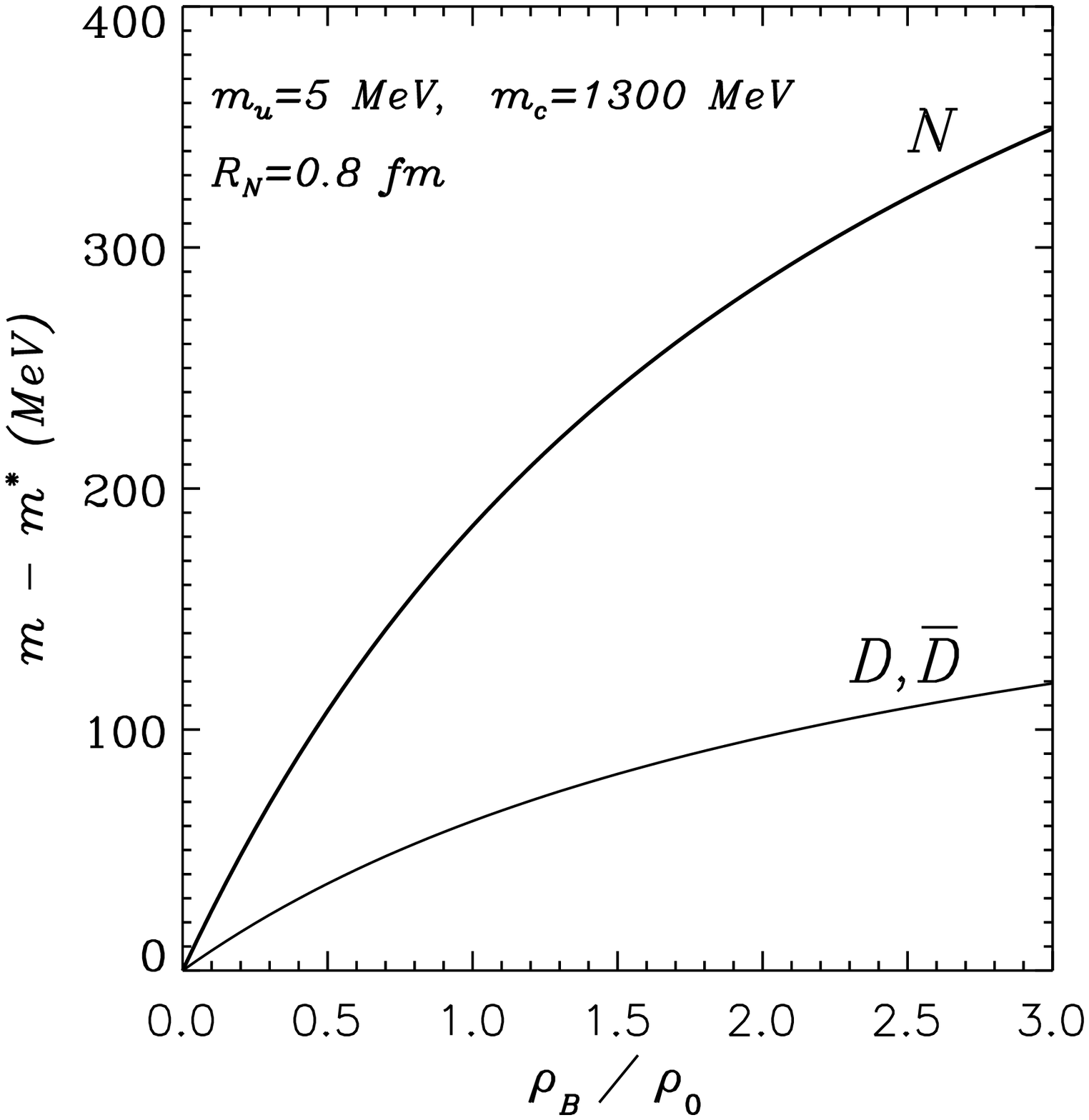,height=7cm}
\vspace{-0.5cm}
\caption{Effective masses of the nucleon, physical $\eta$, physical $\omega$,  
$D$ and $\Dbar$ mesons, and those calculated based on SU(3) quark model basis
(the dotted lines), which are given in Eqs.~(\protect\ref{mixing1}) 
-~(\protect\ref{mixing3}).
The two cases for the $\omega$ meson are almost degenerate.
(Normal nuclear matter density, $\rho_0$, is 0.15 fm$^{-3}$.)}
\label{mesonmatter}
\end{center}
\end{figure}

First, we show in Fig.~\ref{mesonmatter} 
the mass shifts of the $\eta, \omega$ and $D\, (\Dbar)$ mesons, 
calculated in symmetric nuclear matter~\cite{etao,dmeson}. 
The masses for the $\eta$ and $\omega$ are calculated using the 
physical states, i.e., the superpositions of 
the octet and singlet states with the mixing angles, 
$\theta_P$ for the $\eta - \eta'$ and $\theta_V$
for the $\phi - \pomega$~\cite{pdata}:
\bg
\xi  &=& \xi_8 \cos\theta_{P,V} - \xi_1 \sin\theta_{P,V},\quad
\xi' = \xi_8 \sin\theta_{P,V} + \xi_1 \cos\theta_{P,V},
\label{mixing1}\\
{\rm with}\hspace{1cm} & &\nn
\\
\xi_1 &=& \frac{1}{\sqrt{3}}\; (u\ubar + d\dbar + s\sbar),\quad
\xi_8 = \frac{1}{\sqrt{6}}\; (u\ubar + d\dbar - 2 s\sbar),
\label{mixing2}
\\
(\xi, \xi') &=& (\eta, \eta')\quad {\rm or}\quad (\phi, \omega),
\label{mixing3}
\en
where the values, $\theta_P = - 10^\circ$ and
$\theta_V = 39^\circ$~\cite{pdata} are used.

One can easily see that the effect of the singlet-octet mixing
is negligible for the $\omega$ mass in matter,
whereas it is important for the $\eta$ mass.
In addition, $(m_{D,\bar{D}} - m^*_{D,\bar{D}}) \simeq
\frac{1}{3} (m_N - m^*_N)$ is well realized as is
expected~\cite{finite2,hyper}.

%
%
\begin{figure}[hbt]
\begin{center}
\hspace*{-1cm}
\epsfig{file=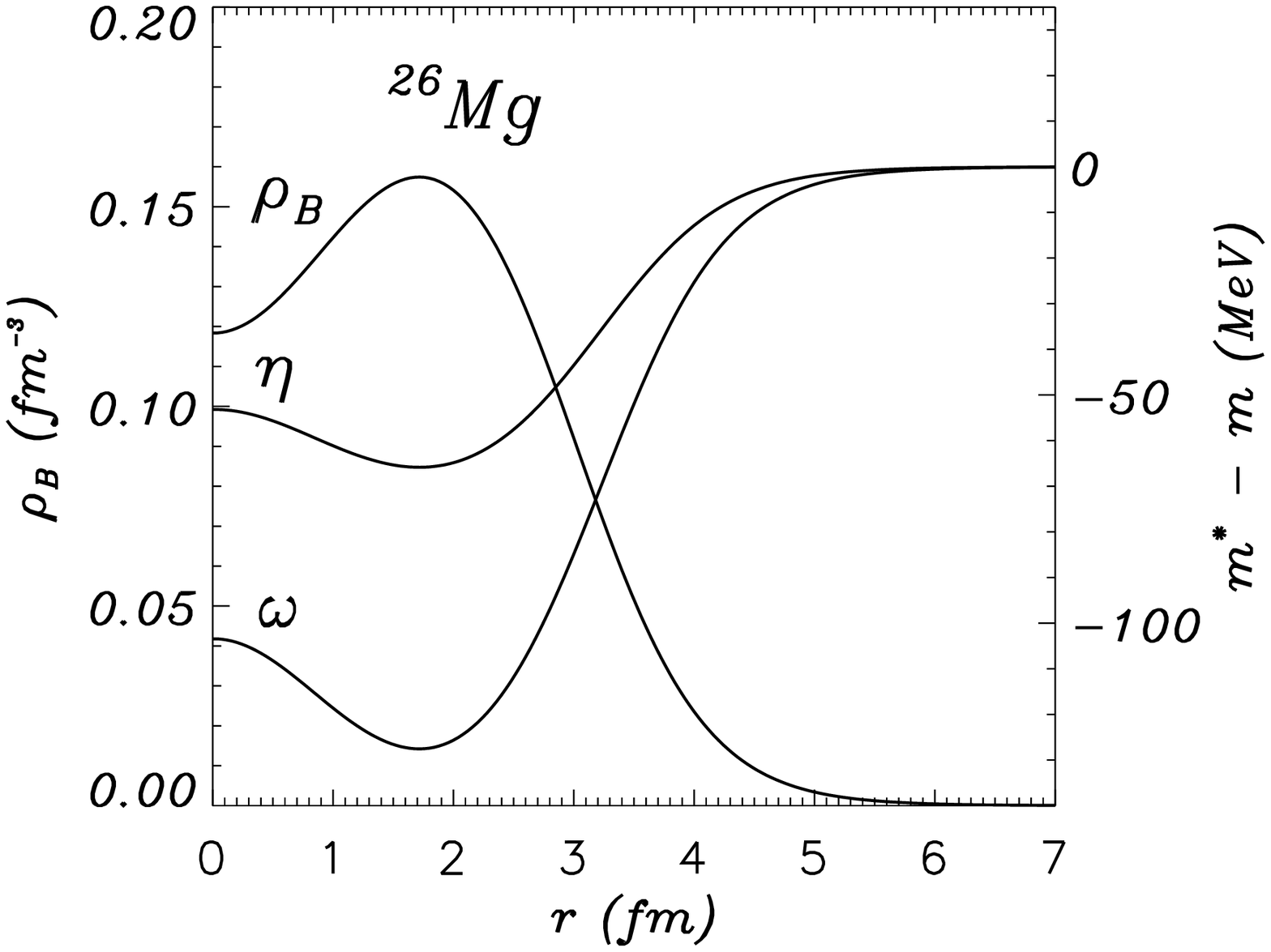,height=7cm}\quad
\epsfig{file=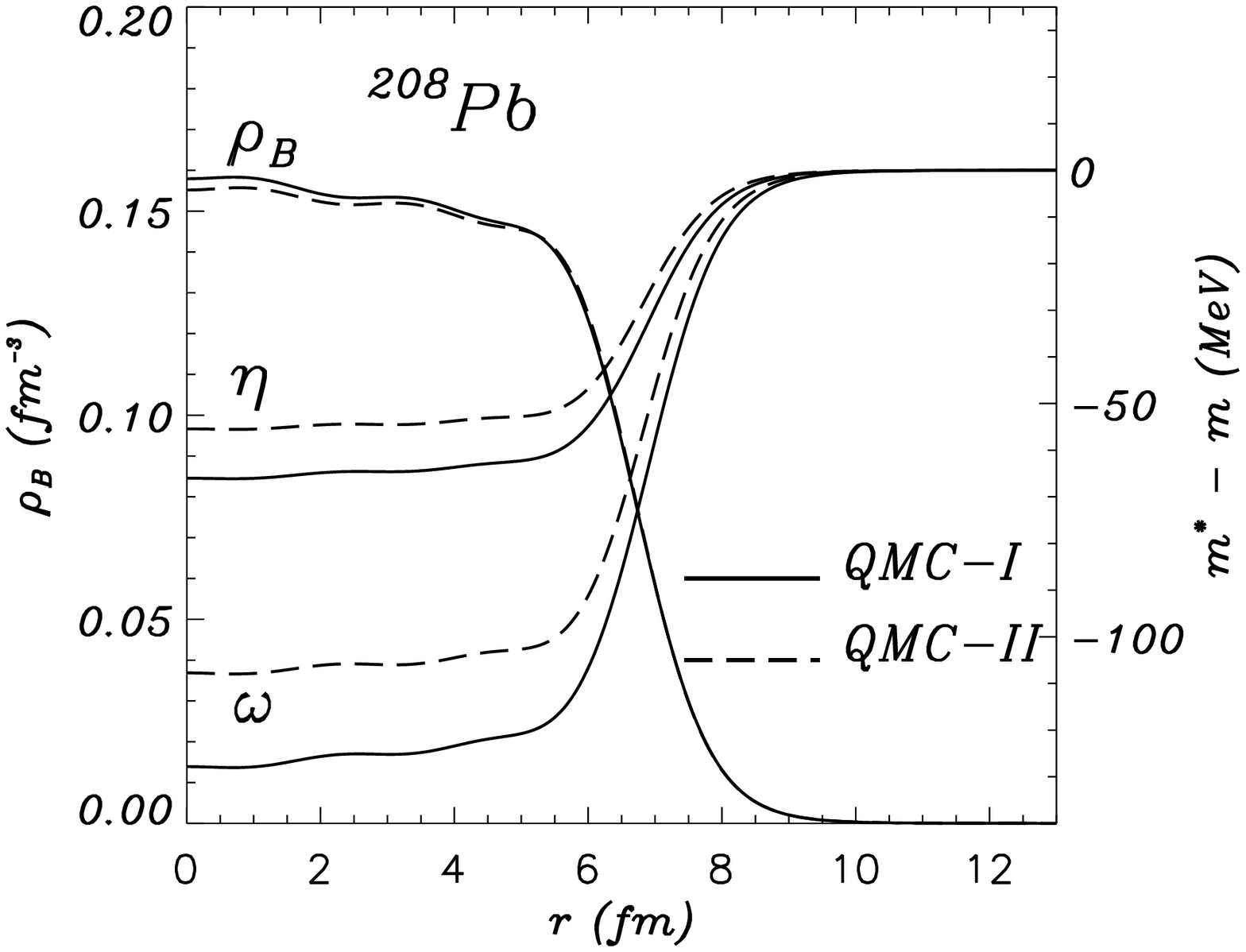,height=7cm}
\vspace{-0.5cm}
\caption{Potentials for the $\eta$ and $\omega$ mesons,
($m^*_\eta(r) - m_\eta$) and ($m^*_\pomega(r) - m_\pomega$), calculated
in QMC-I~\protect\cite{finite1} for $^{26}$Mg and $^{208}$Pb.
For $^{208}$Pb the potentials are also shown for
QMC-II~\protect\cite{finite2}.}
\label{etaopot}
\end{center}
\end{figure}
%
\begin{figure}[hbt]
\begin{center}
\hspace*{-1cm}
\epsfig{file=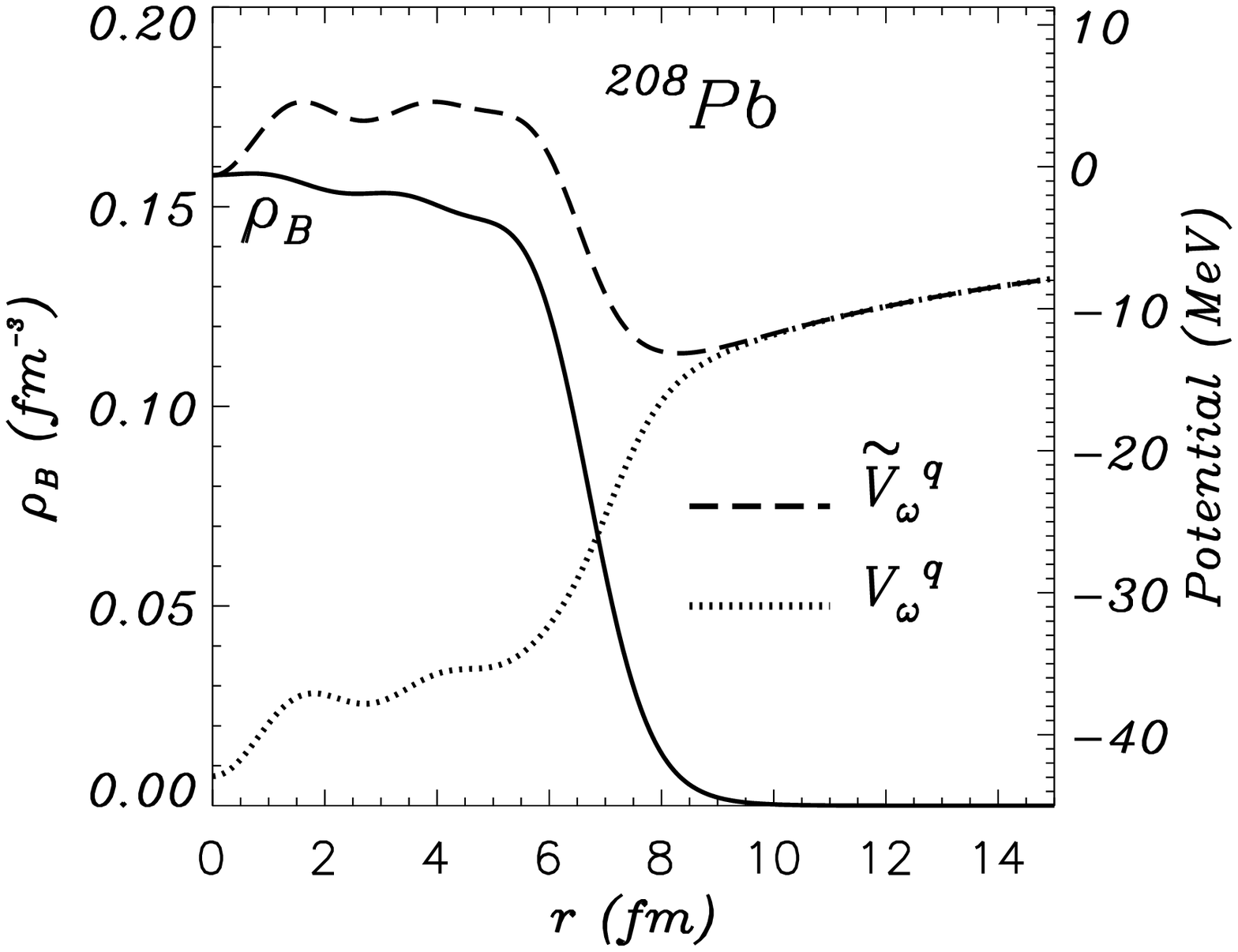,height=7cm}\quad
\epsfig{file=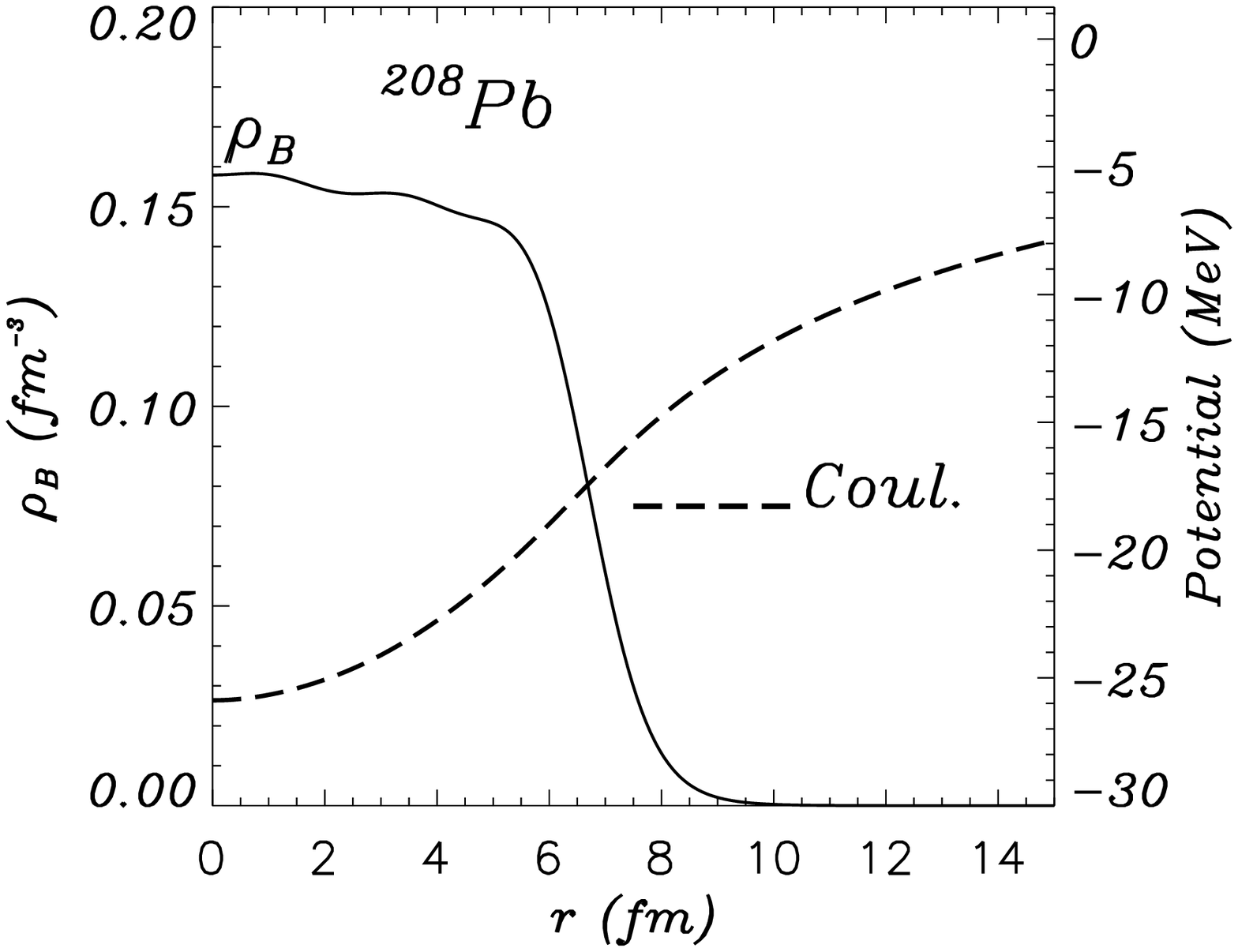,height=7cm}
\vspace{-0.5cm}
\caption{The left panel shows sum of the scalar,
vector and Coulomb potentials for the $D^-$
meson in $^{208}$Pb for two cases,
$(m^*_{D^-}(r) - m_{D^-}) + \tilde{V}^q_\omega(r)
+ \frac{1}{2} V^q_\rho(r) - A(r)$ (the dashed line) and
$(m^*_{D^-}(r) - m_{D^-}) + V^q_\omega(r)
+ \frac{1}{2} V^q_\rho(r) - A(r)$ (the dotted line),
where $\tilde{V}^q_\omega(r) = 1.4^2 V^q_\omega(r)$. The right
panel shows the Coulomb potential.
}
\label{dmespot}
\end{center}
\end{figure}
%

Next, we consider potentials for the mesons in a nucleus. 
A detailed description of the Lagrangian density and the
mean-field equations of motion needed to describe a finite nucleus
is given in Refs.~\cite{finite,finite1}. At position $\vr$ in the 
nucleus (the coordinate origin is taken at the center of the nucleus),
the Dirac equations for the quarks and antiquarks in the 
meson bags ($|\vx - \vr| \le$ bag radius)
are given by~\cite{etao,dmeson,kaon}:
\bg
\left[ i \gamma \cdot \partial_x - (m_q - V^q_\sigma(\vr))
\mp \gamma^0
\left( V^q_\omega(\vr) + \frac{1}{2} V^q_\rho(\vr) \right) \right]
\left(\begin{array}{c} \psi_u(x)\\ \psi_\ubar(x)\\ \end{array}\right)
 &=& 0,
\label{diracu}
\\
\left[ i \gamma \cdot \partial_x - (m_q - V^q_\sigma(\vr))
\mp \gamma^0
\left( V^q_\omega(\vr) - \frac{1}{2} V^q_\rho(\vr) \right) \right]
\left(\begin{array}{c} \psi_d(x)\\ \psi_\dbar(x)\\ \end{array} \right)
 &=& 0,
\label{diracd}
\\
\left[ i \gamma \cdot \partial_x - m_{s,c} \right]
\psi_{s,c} (x)\,\, ({\rm or}\,\, \psi_{\sbar,\cbar}(x)) &=& 0.
\label{diracsc}
\en
The mean-field potentials for a bag centered at position $\vr$ in 
the nucleus, which are approximated to be constants in the entire 
bag volume, are defined by $V^q_\sigma(\vr) = g^q_\sigma
\sigma(\vr), V^q_\omega(\vr) = g^q_\omega \omega(\vr)$ and
$V^q_\rho(\vr) = g^q_\rho b(\vr)$, with $g^q_\sigma, g^q_\omega$ and
$g^q_\rho$ the corresponding quark and meson-field coupling
constants. The mean meson fields, which will depend only on the 
distance, $r = |\vr|$,  are
calculated self-consistently by solving Eqs.~(23) -- (30) of
Ref.~\cite{finite1}.

The corresponding quark eigenenergies in the bag 
in units of $1/R_j^*$ ($j=\eta,\omega,D,\Dbar$) 
are given by~\cite{etao,dmeson,kaon}
\bg
\left( \begin{array}{c} \e_u(\vr) \\ \e_{\ubar}(\vr) \end{array} \right)
&=& \Omega_q^*(\vr) \pm R_j^* \left(
V^q_\omega(\vr) + \frac{1}{2} V^q_\rho(\vr) \right),
\label{uqenergy}
\\
\left( \begin{array}{c} \e_d(\vr) \\ \e_{\dbar}(\vr) \end{array} \right)
&=& \Omega_q^*(\vr) \pm R_j^* \left(
V^q_\omega(\vr) - \frac{1}{2} V^q_\rho(\vr) \right),
\label{dqenergy}
\\
\e_{s,c}(\vr) &=& \e_{\sbar,\cbar}(\vr) = \Omega_{s,c}(\vr),
\label{cenergy}
\en
where $\Omega_q^*(\vr) = \sqrt{x_q^2 + (R_j^* m_q^*)^2}$, with
$m_q^* = m_q - g^q_\sigma \sigma(\vr)$ and
$\Omega_{s,c}(\vr) = \sqrt{x_{s,c}^2 + (R_j^* m_{s,c})^2}$.
The bag eigenfrequencies, $x_q$ and $x_{s,c}$, are
determined by the usual, linear boundary condition~\cite{finite,finite1}.
The meson masses
in the nucleus at position $\vr$, are calculated by
(see also Eqs.~(\ref{mixing1}) -~(\ref{mixing3}) 
for the $\eta$ and $\omega$):
\bg
m_{\eta,\omega}^*(\vr) &=& \frac{2 [a_{P,V}^2\Omega_q^*(\vr)
+ b_{P,V}^2\Omega_s(\vr)] - z_{\eta,\omega}}{R_{\eta,\omega}^*}
+ {4\over 3}\pi R_{\eta,\omega}^{* 3} B,
\label{metao}\\
m_{D,\Dbar}^*(\vr) &=& \frac{\Omega_q^*(\vr)
+ \Omega_c(\vr) - z_{D,\Dbar}}{R_{D,\Dbar}^*}
+ {4\over 3}\pi R_{D,\Dbar}^{* 3} B,
\label{md}\\
& &\left.\frac{\partial m_j^*(\vr)}
{\partial R_j}\right|_{R_j = R_j^*} = 0, \quad\quad 
(j = \eta,\omega,D,\Dbar),
\label{equil}\\
{\rm with}\qquad \hfill\nn \\
a_{P,V} &=& \frac{1}{\sqrt{3}} \cos\theta_{P,V}
- \sqrt{\frac{2}{3}} \sin\theta_{P,V},\quad
b_{P,V} = \sqrt{\frac{2}{3}} \cos\theta_{P,V}
+ \frac{1}{\sqrt{3}} \sin\theta_{P,V}.\qquad
\label{abpv}
\en
In Eqs.~(\ref{metao}), the constants $z_j\, (j=\eta,\omega,D,\Dbar)$  
parameterize the sum of the center-of-mass and gluon fluctuation effects,
which are determined to reproduce the corresponding physical meson masses
in free space. Note that $z_j$ and $B$ are independent of density.

In this study we chose the values,  
$(m_q, m_s, m_c) = (5, 250, 1300)$ MeV for the 
current quark masses, and $R_N = 0.8$
fm for the bag radius of the nucleon in free space. (The other 
input parameters at the quark and hadronic levels   
are given in Refs.~\cite{etao,finite1,dmeson}.)
We stress here that exactly the same coupling constants  
in QMC, $g^q_\sigma$, $g^q_\omega$ and
$g^q_\rho$, are used for the light quarks in the mesons as in the
nucleon. However, in studies of the kaon system, we found that it was
phenomenologically necessary to increase the strength of the vector
coupling to the non-strange quarks in the $K^+$ 
($g^q_\omega \to 1.4^2 g^q_\omega$) 
in order to reproduce the empirically extracted $K^+$-nucleus
interaction~\cite{kaon}. It is not yet clear whether this is a
specific property of the $K^+$, which is a pseudo-Goldstone boson, or
a general feature of the interaction of a light quark. Thus, we show 
results for the $\Dbar$ bound state energies with both choices for 
this potential, in order to test the theoretical uncertainty. 
For the larger $\omega$ meson coupling, suggested by $K^+A$ scattering,
$V^q_\omega(r)$ is replaced by $\tilde{V}^q_\omega(r) = 1.4^2 V^q_\omega(r)$.
Note that the $\rho$ meson mean field potential, $V^q_\rho(r)$, is negative
in a nucleus with a neutron excess, such as e.g., $^{208}$Pb.

In Figs.~\ref{etaopot} and~\ref{dmespot} we show the calculated
potentials, for the $\eta$ and $\omega$, and the
$D^-$ mesons, respectively. For the $D^-$ we show also the 
Coulomb potential which ensures the formation of atomic bound states. 
The left panel in Fig.~\ref{dmespot} shows the {\it naive} sum of the
scalar and vector potentials for the $D^-$, for the two choices of
the vector potentials, $\tilde{V}^q_\omega(r)$ (the dashed line)
and ${V}^q_\omega(r)$ (the dotted line), while the right panel shows 
the Coulomb potential.

Because the $D^-$ meson is heavy and may be described well
in the (nonrelativistic) Schr\"{o}dinger equation, one
expects the existence of the $_{D^-}^{208}$Pb bound states
just from inspection of the {\it naive} sum of the potentials,
in a way which does not distinguish the Lorentz vector or
scalar character.

\section{Results}

\begin{table}[htb]
\begin{center}
\caption{
Calculated $\eta$ and $\omega$ meson bound state energies (in MeV),
$E_j = Re (E^*_j)\,(j=\eta,\omega)$,
in QMC~\protect\cite{etao} and those of the $\omega$ in
the Walecka model with $\sigma$-$\omega$ mixing effect~\protect\cite{qhdo}.
The complex eigenenergies are given by,
$E^*_j = E_j + m_j - i \Gamma_j/2$. 
}
\label{etaoenergy}
\begin{tabular}[t]{lc|cc|cc|cc}
\hline \hline
& &$\bm \gamma_\eta=0.5$ &(QMC) &$\bm \gamma_\omega$=0.2
&(QMC) &$\bm \gamma_\omega=0.2$ &(QHD)\\
\hline \hline
& &$E_\eta$ &$\Gamma_\eta$ &$E_\omega$ &$\Gamma_\omega$
&$E_\omega$ &$\Gamma_\omega$\\
\hline
$^{6}_j$He &1s &-10.7&14.5 &-55.6&24.7 &-97.4&33.5 \\
\hline
$^{11}_j$B &1s &-24.5&22.8 &-80.8&28.8 &-129&38.5 \\
\hline
$^{26}_j$Mg &1s &-38.8&28.5 &-99.7&31.1 &-144&39.8 \\
            &1p &-17.8&23.1 &-78.5&29.4 &-121&37.8 \\
            &2s & --- & --- &-42.8&24.8 &-80.7&33.2 \\
\hline \hline
$^{16}_j$O &1s &-32.6&26.7 &-93.4&30.6 &-134&38.7 \\
           &1p &-7.72&18.3 &-64.7&27.8 &-103&35.5 \\
\hline
$^{40}_j$Ca &1s &-46.0&31.7 &-111&33.1  &-148&40.1 \\
            &1p &-26.8&26.8 &-90.8&31.0 &-129&38.3 \\
            &2s &-4.61&17.7 &-65.5&28.9 &-99.8&35.6  \\
\hline
$^{90}_j$Zr &1s &-52.9&33.2 &-117&33.4  &-154&40.6 \\
            &1p &-40.0&30.5 &-105&32.3  &-143&39.8 \\
            &2s &-21.7&26.1 &-86.4&30.7 &-123&38.0 \\
\hline
$^{208}_j$Pb &1s &-56.3&33.2 &-118&33.1 &-157&40.8 \\
             &1p &-48.3&31.8 &-111&32.5 &-151&40.5 \\
             &2s &-35.9&29.6 &-100&31.7 &-139&39.5 \\
\hline \hline
\end{tabular}
\end{center}
\end{table}
%
%
\begin{figure}[hbt]
\begin{center}
\epsfig{file=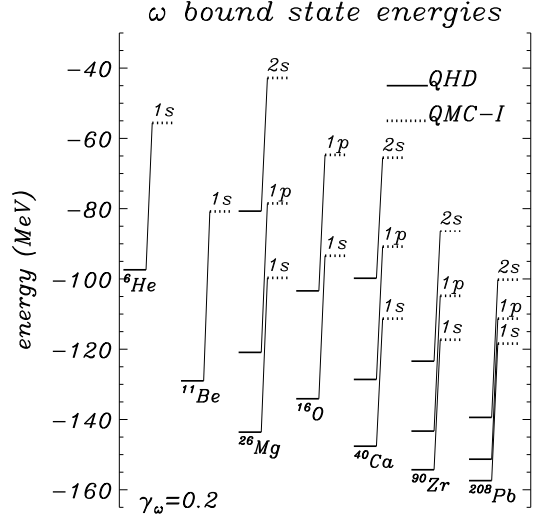,height=8cm}
\vspace{-1cm}
\caption{Calculated $\omega$ meson bound state energy levels in
QHD and QMC. Detailed data for QMC can be seen in Ref.\protect\cite{etao}.
}
\label{speomega}
\end{center}
\end{figure}
%
\begin{table}[htb]
\begin{center}
\caption{
Calculated $D^-$, $\d0bar$ and $D^0$ meson bound state energies (in MeV)
in $^{208}$Pb for different potentials.
The widths for the mesons are all set to zero, both
in free space and inside $^{208}$Pb. 
Note that the $D^0$ bound states
energies calculated with $\tilde{V}^q_\omega$ will be much larger than
those calculated with $V^q_\omega$ (in absolute value).
}
\label{denergy}
\vspace{0.2cm}
\begin{tabular}[t]{l|ccc|ccc}
state  &$D^- (\tilde{V}^q_\omega)$ &$D^- (V^q_\omega)$
&$D^- (V^q_\omega$, no Coulomb) &$\d0bar (\tilde{V}^q_\omega)$
&$\d0bar (V^q_\omega)$ &$D^0 (V^q_\omega)$ \\
\hline
                         1s &-10.6 &-35.2 &-11.2 &unbound &-25.4 &-96.2\\
                         1p &-10.2 &-32.1 &-10.0 &unbound &-23.1 &-93.0\\
                         2s & -7.7 &-30.0 & -6.6 &unbound &-19.7 &-88.5\\
\hline 
\end{tabular}
\end{center}
\end{table}
%

Here we calculate the bound state energies for the
mesons using the mean field potentials obtained in QMC.
We consider the situation of almost zero momentum for the mesons. 
Then, after imposing the Lorentz condition, solving the Proca equation  
becomes equivalent to solving the Klein-Gordon equation, because the  
transverse and longitudinal masses for the vector meson are degenerate.
Thus, we may solve the following form of the Klein-Gordon equation  
for the $\eta$, $\omega$, $D$ and $\Dbar$mesons:
\bg
[ \nabla^2 &+& (E_j - V^j_v(r))^2- \tilde{m}^{*2}_j(r) ]\,
\phi_j(\vr) = 0,\qquad (j=\eta,\omega,D,\Dbar), 
\label{kgequation}
\\
{\rm with}\hspace{1cm}& & \nn
\\
\tilde{m}^*_j(r) &=& m^*_j(r) - \frac{i}{2}
\left[ (m_j - m^*_j(r)) \gamma_j + \Gamma_j \right]\,
\equiv\, m^*_j(r) - \frac{i}{2} \Gamma^*_j (r),
\label{width}
\en
where $E_j$ is the total energy of the meson 
(the binding energy is $E_j-m_j$), $V^j_v(r)$, $m_j$ and $\Gamma_j$
are the sum of the vector and Coulomb potentials,
the corresponding masses and widths
in free space~\cite{pdata,etao,dmeson}, and 
$\gamma_j$ are treated as phenomenological
parameters to describe the in-medium meson widths,
$\Gamma^*_j(r)$. 
According to the estimates in Refs.~\cite{hayano,fri},
the widths of the mesons in nuclei and at normal nuclear matter density
are, $\Gamma^*_\eta \sim 30 - 70$ MeV~\cite{hayano}
and $\Gamma^*_\pomega \sim 30 - 40$ MeV~\cite{fri}, respectively.
Thus, we show the bound state energies calculated for the values
of the parameters, $\gamma_\eta = 0.5$ and $\gamma_\omega = 0.2$, 
which are expected to correspond best with experiment. 
For the $D^-$ and $\d0bar$, the widths are set to zero which is exact, 
whereas those for the $D^0$ does not make sense. Nevertheless,  
some results for the $D^0$ calculated using the zero width will 
be listed for the purpose of comparison. 
Eq.~(\ref{kgequation}) is  
solved in momentum space~\cite{landau}, where 
extra care is taken for the treatment of the long range 
Coulomb potential for the $D^-$ (see Refs.~\cite{dmeson,landau}). 

The calculated  meson-nucleus bound state energies
are listed in Tables~\ref{etaoenergy} and~\ref{denergy}.
Results for the $\omega$-meson are also shown for  
the Walecka model~\cite{qhdo}.

\section{Summary}

We have investigated several possible meson-nucleus bound states using QMC.
Our results suggest that $\eta$ and $\omega$ mesons should be 
bound in all the nuclei considered. Furthermore, the $D^-$ meson
should inevitably be bound in $^{208}$Pb, due to two quite
different mechanisms, namely, the scalar and attractive
$\sigma$ mean field, with the assistance of the Coulomb force.
The existence of any bound states at all would give us important 
information concerning the role of the Lorentz scalar $\sigma$ field,  
and hence dynamical symmetry breaking; in other words, 
information on the partial restoration of chiral symmetry in medium.
\\

\noindent
{\bf Ackowledgement}\\
The author would like to thank D.H. Lu, K. Saito and 
A.W. Thomas for exciting collaborations. The results reported here
are based on the work investigated together with them.
This work was supported by the Australian Research Council.
%

\end{document}